\documentclass[iop]{emulateapj}
\usepackage{graphicx}
\usepackage{latexsym}
\usepackage{amssymb}
\usepackage{longtable}
\usepackage{amsmath}
\usepackage{url}
\def\kms{km s$^{-1}$ }
\def\Msun{\ifmmode {\rm\,\it{M_\odot}}\else ${\rm\,M_\odot}$\fi}
\def\Rsun{\ifmmode {\rm\,\it{R_\odot}}\else ${\rm\,R_\odot}$\fi}
\def\simlt{\mathrel{\spose{\lower 3pt\hbox{$\mathchar"218$}}
     \raise 2.0pt\hbox{$\mathchar"13C$}}}
\def\simgt{\mathrel{\spose{\lower 3pt\hbox{$\mathchar"218$}}
\     \raise 2.0pt\hbox{$\mathchar"13E$}}}

\usepackage[pdftex,backref,breaklinks,colorlinks,citecolor=blue]{hyperref}
\usepackage[all]{hypcap}

\begin{document}

\title{Investigating the origin of hot gas lines in Herbig Ae/Be Stars}
\author{P. Wilson Cauley}
\email{pcauley@wesleyan.edu}
\affil{Wesleyan University}
\affil{Department of Astronomy and Van Vleck Observatory, 96 Foss Hill Dr., Middletown, CT 06459}
\author{Christopher M. Johns--Krull}
\email{cmj@rice.edu}
\affil{Rice University}
\affil{Department of Physics and Astronomy, 6100 Main St., MS 108, Houston, TX 77005}

\begin{abstract}
We analyze high--resolution UV spectra of a small sample of Herbig Ae/Be stars (HAEBES) in order
to explore the origin of the $T\sim10^5$ K gas in these stars. The \ion{C}{4}
1548,1550 \AA\hspace{0pt} line luminosities are compared to non--simultaneous accretion rate estimates for
the objects showing \ion{C}{4} emission. We show that the correlation between $L_{CIV}$ and $\dot{M}$
previously established for classical T Tauri stars (CTTSs) seems to extend into the HAEBE mass
regime, although the large spread in literature $\dot{M}$ and $A_V$ values makes the actual
relationship highly uncertain. With the
exception of DX Cha, we find no evidence for hot, optically thick winds in our HAEBE sample. All other objects showing
clear doublet emission in \ion{C}{4} can be well described by a two component (i.e., a single component for each doublet member)
or four component (i.e., two components for each doublet member) Gaussian emission line fit. The
morphologies and peak-flux velocities of these lines suggest they are formed in weak, optically thin stellar winds and not in an
accretion flow, as is the case for the hot lines observed in CTTSs. The lack of strong outflow signatures and lack
of evidence for line formation in accretion flows is consistent with the conclusion presented in our recent
optical and \ion{He}{1} 10830 \AA\hspace{0pt} studies that the immediate circumstellar environments of
HAEBES, in general, are not scaled--up analogs of the immediate environments around CTTSs.
The conclusions presented here for hot gas lines around HAEBES should be verified with a larger sample of
objects. 
\end{abstract}

\newpage

\section{INTRODUCTION}
\label{sec:sec1}

Mass loss from young stars, in the form of stellar and disk winds, is a critical evolutionary
process that largely determines the rotation period of the star when it reaches the main sequence.
Mass loss also removes material from the disk that would otherwise be available for the formation of
planets. Understanding the physical properties of these winds and how they are launched is thus
crucial to building a complete understanding of early stellar and planetary evolution. 

Stellar and disk winds are intimately tied to the accretion process
\citep{hartigan95,matt05,roma09,zanni13}. There is evidence that magnetospheric accretion, which is
the dominant accretion mechanism for classical T Tauri stars (CTTSs) and perhaps many Herbig Ae stars
\citep{muzz01,muzz04,cauley14}, is able to drive both stellar winds from near the surface of the
star and magneto-centrifugal winds from near the star-disk interaction region
\citep{edwards06,cauley14}. Boundary layer accretion, which may operate more prevalently in the
higher mass Herbig Be stars \citep{mottram07,cauley14}, is also capable of driving strong outflows from near
the surface of rapidly rotating stars \citep{shu88}. The physical conditions and energetics of the
accretion flow thus have a strong impact on the nature of the outflow. By examining spectral
diagnostics that probe different temperature and density regimes, estimates of the physical
conditions in the mass flows of young stars can be obtained, which in turn could give additional clues to the driving
mechanism.

Numerous studies, both observational and theoretical, have investigated the physical conditions of
accreting material and outflows around both CTTSs and Herbig Ae/Be stars (HAEBES)
\citep[e.g.,][]{catala88,calvet98,muzz04,kurosawa,edwards13}. In general, CTTSs show evidence of
magnetically dominated accretion flows with temperatures and densities of $\sim$10$^4$ K and
$\sim$10$^{11}$ cm$^{-3}$, respectively, along the length of the funnel flows and densities $\sim$10 times
higher in the immediate pre-shock gas near the stellar surface. There is a large spread in the inferred temperatures
and the inferred densities increase for objects with larger accretion rates \citep[e.g.,][]{muzz98,ingleby13,edwards13}. 
Simulations of stellar winds and inner disk winds
require similar physical characteristics to roughly reproduce observed line profiles
\citep{kurosawa}. We note that the physical mechanism responsible for accelerating accretion--driven
stellar winds is still unknown. Similar temperature and density ranges in magnetic accretion flows
onto HAEBES have been successful at modeling observed Balmer and \ion{Na}{1} line profiles \citep{muzz04}.

Outflow signatures (i.e., strong blue--shifted absorption) have been observed in high--temperature
lines (e.g., \ion{C}{4}) in a handful of HAEBES \citep{praderie,grady96}. For the HAEBES AB Aur, BD+46$^\circ$3471,
HD 250550, and BD+61$^\circ$154, modeling
the \ion{C}{4} line formation in an expanding chromosphere (i.e., the base of the stellar wind) results in wind temperatures of $\sim$15,000--20,000
K and mass loss rates of $\sim5\times10^{-8}$ \Msun\hspace{0pt} yr$^{-1}$ \citep{catala84,catala88,bouret}.
Temperatures of order $\sim$10$^5$ K in regions above the photosphere have also been invoked to  
explain observed \ion{N}{5} line profiles in AB Aur \citep{bouret97}.
Strong \textit{stellar} wind signatures are also seen in the moderately high--temperature ($T\sim20,000$ K) He I $\lambda$10830 line in a
large fraction ($\sim$40\%) of HAEBES \citep{cauley14}. Due to the large photoionizing flux of the
model HAEBE chromospheres, in addition to X--ray and UV emission from accretion flows, the highest temperature 
spectral lines, such as \ion{C}{4} and \ion{N}{5}, do not necessarily require temperatures of $\sim$10$^5$ K to form \citep{catala88,kwan11}.
Instead, photoionization allows these lines to form at much lower temperatures ($\sim$15,000-20,000 K; \citet{catala84,catala88}), suggesting they are
likely representative of cooler regions at the base of the outflow.

Although HAEBES do not have fully convective envelopes, and therefore
are not expected to sustain active chromospheres and coronae, there are indications that
they do indeed host extended atmospheres responsible for some of the observed high-temperature
emission. \cite{bouret} find that a dense, expanding chromosphere is required to adequately describe
the blue-shifted absorption signatures in the UV lines of AB Aur, HD 250550, BD+46$^\circ$3471, and 
BD+61$^\circ$154. In an attempt to explain X-ray emission observed in HAEBES by \citet{zinnecker}, 
\citet{tout95} suggested that shear layers in the rapidly rotating atmospheres of HAEBES are
responsible for driving a dynamo that is capable of sustaining a magnetic field and generating
the observed activity levels. Another possibility for magnetic field generation in HAEBES is a
near-surface convection layer produced by the ionization of hydrogen and helium in A-type stars
and iron in the hotter B-type stars \citep{cantiello11,drake14}. 
Although magnetic fields have only been detected on a handful of
HAEBES \citep[e.g., see][]{alecian13}, weak small-scale fields may be below current detection
limits, preventing the \citet{tout95} scenario from being ruled out for HAEBES in general. More
recent detections of X-ray emission from HAEBES are consistent with the conclusion that at least
some these objects are intrinsic X-ray sources, although the exact fraction of intrinsic X-ray emitters
compared to those with unresolved low mass companions is uncertain \citep{stelzer06,stelzer09,guenther09}. However,
it is still unclear whether the intrinsic X-ray emission is generated by the star itself, whether
it is powered by an accretion flow or stellar wind \citep[e.g., ][]{guenther09,drake14}, or a
combination of both. Although some evidence points to their existence, more investigation is 
needed into the nature of extended atmospheres around HAEBES.

A recent study by \defcitealias{ardila}{A13}\citet[][ hereafter A13]{ardila} of a large number of
CTTSs shows a complete absence of absorption signatures in high--temperature UV lines. They find
that the observed emission in the resonance UV doublets \ion{N}{5}, \ion{Si}{4}, and \ion{C}{4} can
be well described by a two--component Gaussian fit and are consistent with formation in the
accretion flow and not in a stellar or disk wind. The one HAEBE in their sample, DX Cha, shows
strong evidence of blue--shifted absorption in \ion{C}{4} and \ion{Si}{4}.  The lack of hot wind
signatures observed in CTTSs\footnote{Detection of a 10$^5$ K wind has been reported for TW Hya by
\citet{dupree05,dupree14}, although the original interpretation of the line profiles was contested
by \citet{johnskrull07}.} and the relatively higher rates observed in HAEBES
\citep[e.g.,][]{grady96} suggests that the formation conditions of outflows in the two groups
differs significantly. However, better statistics on the occurrence of hot outflows in HAEBES need
to be established in order to reach more general conclusions concerning differences in the physical
characteristics and the potential launching mechanisms of the outflows in each group.  

Here we present new high spectral resolution observations of UV lines in a sample of 10 HAEBES in
order to place further constraints on the incidence of high temperature outflow and accretion
signatures in HAEBES. The HAEBES in our sample were chosen based on a lack of direct evidence (i.e.,
sub-continuum blue-shifted absorption) for outflows in their optical spectra \citep{cauley15} but
which do show some evidence of mass accretion.\footnote{The possible exceptions are HD 139614 and
DX Cha, which are not examined in \citet{cauley15}, and XY Per. \citet{grady04} report a detection
of a bipolar jet for DX Cha. \citet{mora} fit multiple Gaussians to the emission line profiles of
XY Per, some of which show blue--shifted centroids. However, no sub--continuum blue--shifted
absorption is present.} Evidence of winds in the higher temperature UV lines would thus clarify
whether or not these objects truly lack any significant outflows or if the physical conditions in
the wind render them visible only in the high temperature lines. The most prominent line in our sample is the \ion{C}{4}
1548.2,1550.8 \AA\ doublet. Although we also extract and present the line profiles of \ion{N}{5}
1238.8,1242.8 \AA, \ion{Si}{4} 1398.8,1402.8 \AA, and \ion{He}{2} 1640.4 \AA, these lines are
strongly contaminated by other nearby lines or, in the case of \ion{He}{2}, are weak or not
detectable in most of the targets. Thus we will focus on the morphologies of the \ion{C}{4} doublets
and the physical processes responsible for them. To this end we employ both multi--component
Gaussian fitting and a simple stellar wind model in order to explore whether self-absorption in a
wind is definitively present or whether the line morphologies result from pure emission. In
\autoref{sec:sec2} we present our observations and data reduction procedures. The line profiles and
their properties are briefly discussed in \autoref{sec:sec3}. The Gaussian fitting and stellar wind
model are presented in \autoref{sec:sec4}. A discussion of our results is presented in
\autoref{sec:sec5} and a summary is given in \autoref{sec:sec6}.   
    
\section{OBSERVATIONS AND DATA REDUCTION}
\label{sec:sec2}

The new Hubble Space Telescope (HST) observations were obtained as part of GO-12996 (PI:Johns-Krull) and
are listed in \autoref{tab:tab1}. The DX Cha observations were not a part of this program and are described in 
\citetalias{ardila}. We used the E140M grating with the Space Telescope Imaging Spectrograph (STIS) to observe 5 of the 10 objects.  The
resolving power of these observations is $\sim$45,000, or $\Delta$$v\sim$7 km s$^{-1}$, and they
cover the wavelength region 1140--1710 \AA. The remaining 5 objects were observed with the Cosmic
Origins Spectrograph (COS) G160M and G130M gratings in order to provide wavelength coverage from
$\sim$1225 \AA\hspace{0pt} to 1760 \AA. Small gaps in wavelength coverage exist for the COS data at
$\sim$1280--1290 \AA\hspace{0pt} and 1560--1575 \AA\hspace{0pt} due to the 9 mm gap between detector
elements.  The COS observations have a resolving power of $\sim$20,000, or a velocity resolution of
$\Delta$$v$$\sim$15 km s$^{-1}$. 

\begin{figure*}[htbp]
   \centering
   \includegraphics[scale=.90,trim=30mm 80mm 30mm 10mm]{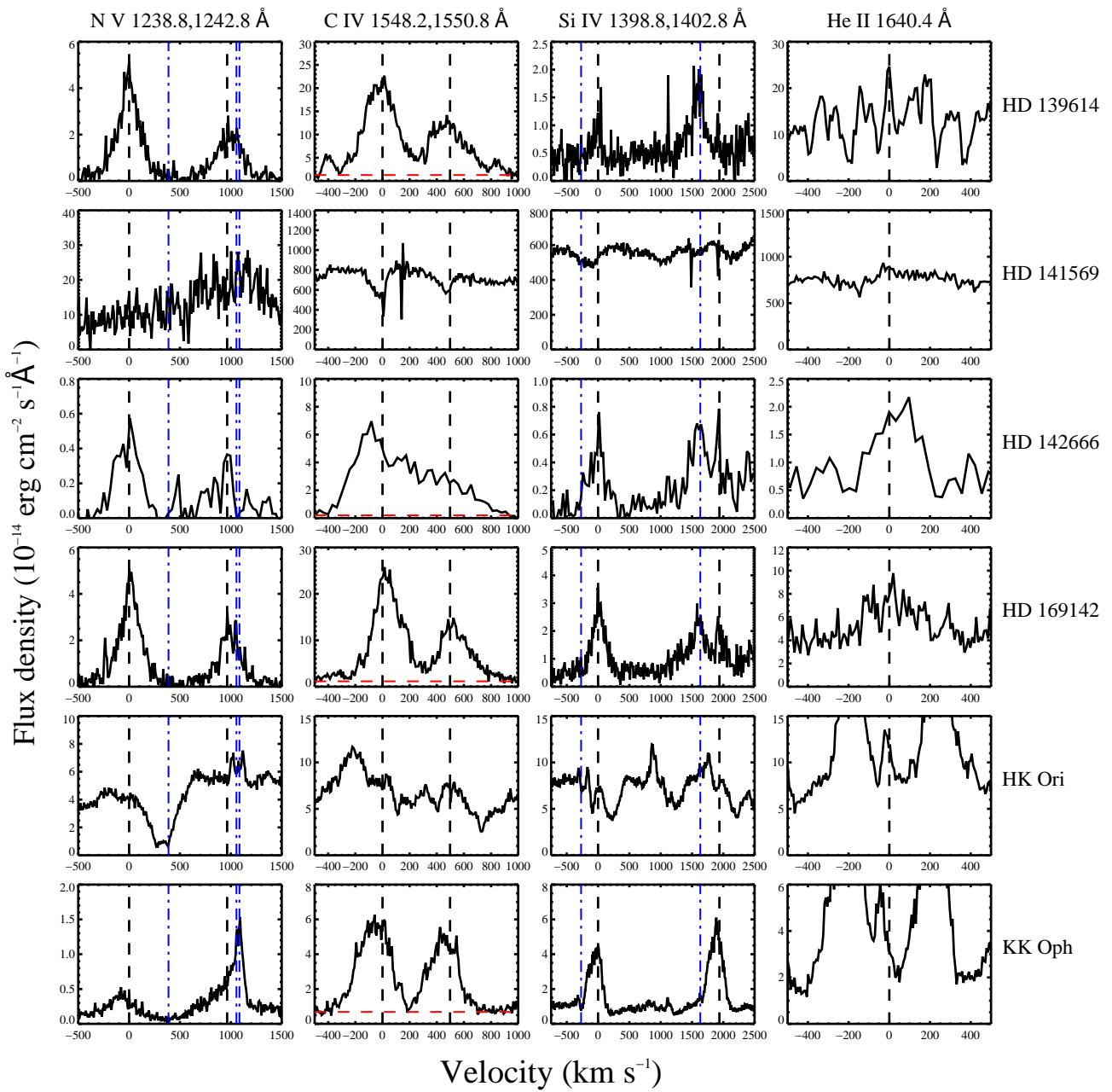} % requires the graphicx package
   \caption{Extracted line profiles for our sample. The rest wavelengths of the line is marked with a vertical dashed black line.
   Rest wavelengths of contaminating features are marked with vertical blue dot--dashed lines. The contaminating lines
   in the \ion{Si}{4} column are the CO A--X (5--0) band head at $-271$ km s$^{-1}$ and \ion{O}{4} $\lambda$1401 at $+1633$
   km s$^{-1}$. In the \ion{N}{5} column the contaminating lines are \ion{Mg}{2} 
   $\lambda$1240.4 at $+387$ km s$^{-1}$ and \ion{N}{1} $\lambda$1243.2,1243.3 at $+1055$ and $+1085$ km s$^{-1}$.
   Continuum flux levels in the \ion{C}{4} panels are marked with a horizontal dashed red line.}
   \label{fig:fig1}
\end{figure*}

\begin{figure*}[htbp]
   \centering
   \includegraphics[scale=.90,trim=30mm 110mm 30mm 10mm]{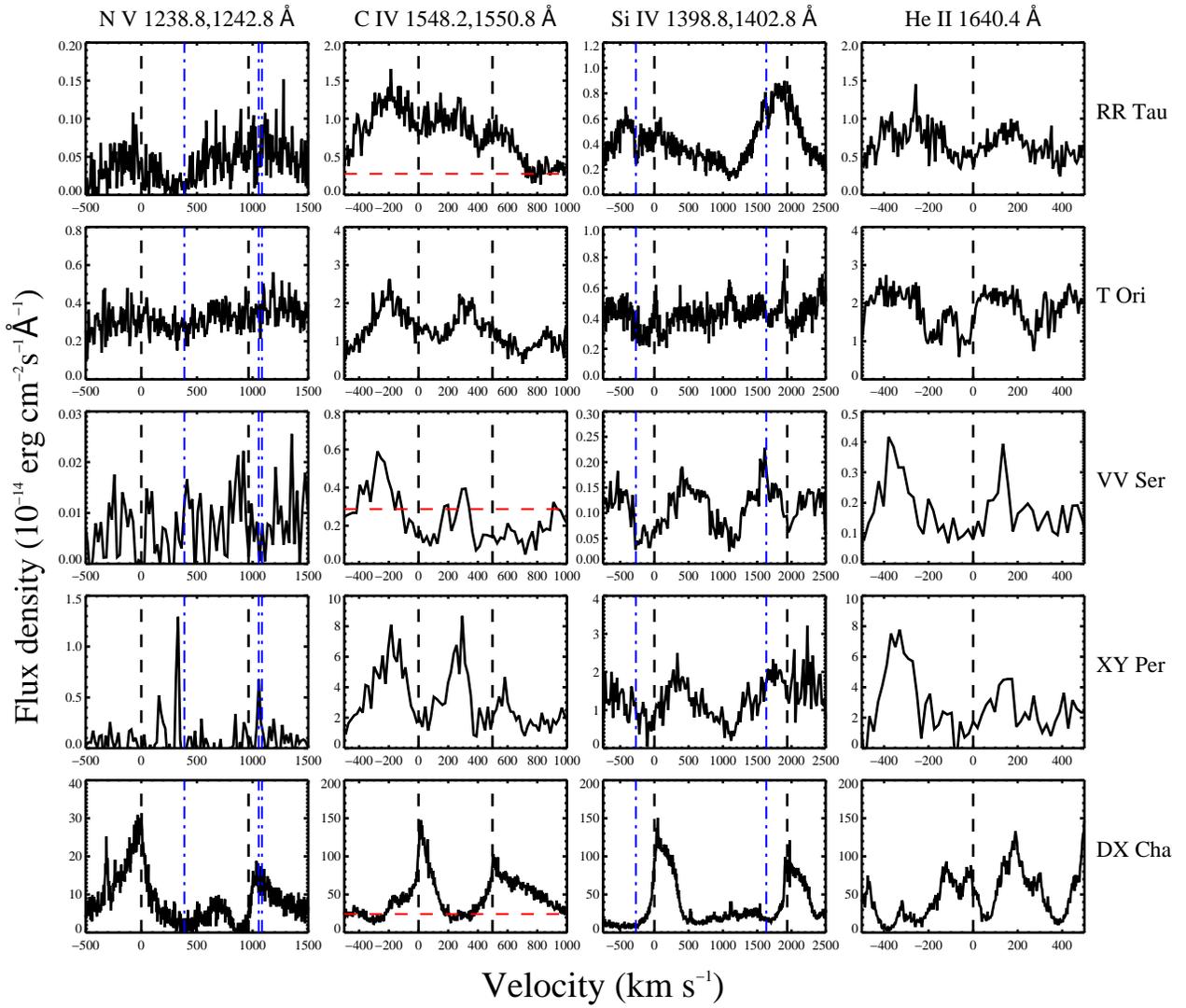} % requires the graphicx package
   \caption{Same as \autoref{fig:fig1} but for the remaining objects in the sample.}
   \label{fig:fig2}
\end{figure*}

All observations were reduced using the latest versions of the \textit{calstis} and \textit{calcos}
pipelines. Pointing data from the raw exposure headers was examined to ensure that objects were
properly acquired and aligned on the instrument apertures. All acquisition and guiding data was
reported as nominal, thus requiring no corrections to the pipeline output.  The STIS echelle orders
from each individual exposure are extracted and combined into a single wavelength and flux array.
Similarly, the spectra from the adjacent COS detector elements are extracted and combined.  Each
STIS object was observed twice on consecutive orbits with the same setup. The spectra from the
consecutive observations were co--added to produce the final spectrum. Radial velocity corrections
are applied to the wavelength arrays using the values given in \autoref{tab:tab2}.
The spectral lines of interest are then individually extracted from the array and continuum normalized. The final line profiles are shown in
\autoref{fig:fig1} and \autoref{fig:fig2}. Due to wavelength overlap of the two COS settings, we
were able to co--add the flux in the \ion{Si}{4} lines from the observations corresponding to the
two different settings. The average signal--to--noise across the binned \ion{C}{4} line profile is
given in column 9 of \autoref{tab:tab1}.

\capstartfalse           
\begin{deluxetable*}{lcccccccc}
%\rotate
%\linespacing{1}
\tablecaption{Log of HST STIS/COS observations \label{tab:tab1}}
\tablewidth{\textwidth}
\tablehead{\colhead{Object ID}&\colhead{Data ID}&\colhead{Instrument}&\colhead{Grating}&
\colhead{Aperture}&\colhead{UT Date}&\colhead{Integration time (s)}&\colhead{Binning (pixels)}&\colhead{S/N @ 1550 \AA}\\
\colhead{(1)}&\colhead{(2)}&\colhead{(3)}&\colhead{(4)}&\colhead{(5)}&\colhead{(6)}&\colhead{(7)}&\colhead{(8)}&\colhead{(9)}}
\tabletypesize{\scriptsize}
\startdata
HD 139614 & OC2H05020 & STIS & E140M & 0.2"x0.2" & 2013--06--24 & 1670 & 3 & 6\\
          & OC2H05030 & STIS & E140M & 0.2"x0.2" & 2013--06--24 & 2758 & &\\
HD 141569 & OC2H04020 & STIS & E140M & 0.2"x0.2" & 2013--07--19 & 1375 & 3 & 25\\
          & OC2H04030 & STIS & E140M & 0.2"x0.2" & 2013--07--19 & 2340& &\\
HD 142666 & OC2H03020 & STIS & E140M & 0.2"x0.2" & 2013--07--16 & 1624 & 10 & 5\\
          & OC2H03030 & STIS & E140M & 0.2"x0.2" & 2013--07--16 & 2763& &\\
HD 169142 & OC2H02020 & STIS & E140M & 0.2"x0.2" & 2013--07--11 & 1732 & 3 & 6\\
          & OC2H02030 & STIS & E140M & 0.2"x0.2" & 2013--07--11 & 2761& &\\
HK Ori    & LC2H06010 & COS  & G160M & PSA       & 2012--12--06 & 1696 & 3 & 15\\
          & LC2H06020 & COS  & G130M & PSA       & 2012--12--06 & 1856& &\\
KK Oph    & LC2H08010 & COS  & G160M & PSA       & 2013--06--18 & 1556 & 3 & 7\\
          & LC2H08020 & COS  & G130M & PSA       & 2013--06--18 & 1856 & &\\
RR Tau    & LC2H07010 & COS  & G160M & PSA       & 2012--12--07 & 1848 & 3 & 5\\
          & LC2H07020 & COS  & G130M & PSA       & 2012--12--07 & 1856& &\\
T Ori     & LC2H09010 & COS  & G160M & PSA       & 2012--12--06 & 1696 & 3 & 7\\
          & LC2H09020 & COS  & G130M & PSA       & 2012--12--06 & 1856 & &\\
VV Ser    & LC2H10010 & COS  & G160M & PSA       & 2012--05--15 & 1696 & 10 & 4\\
          & LC2H10020 & COS  & G130M & PSA       & 2013--05--15 & 1856& &\\
XY Per    & OC2H01020 & STIS & E140M & 0.2"x0.2" & 2012--10--12 & 1613 & 6 & 3\\
          & OC2H01030 & STIS & E140M & 0.2"x0.2" & 2012--10--12 & 2752& &\\
\enddata
\end{deluxetable*}
\capstarttrue

\capstartfalse           
\begin{deluxetable*}{lcccccc}
%\rotate
%\linespacing{1}
\tablecaption{Stellar parameters \label{tab:tab2}}
\tablewidth{\textwidth}
\tablehead{\colhead{}&\colhead{}&\colhead{$v_{rad}$}&\colhead{$M_*$}&\colhead{$R_*$}&\colhead{$v$sin$i$}&\colhead{}\\
\colhead{Object ID}&\colhead{Spectral Type}&\colhead{(km s$^{-1}$)}&
\colhead{($\Msun$)}&\colhead{($\Rsun$)}&\colhead{(km s$^{-1}$)}&\colhead{References$^a$}\\
\colhead{(1)}&\colhead{(2)}&\colhead{(3)}&\colhead{(4)}&\colhead{(5)}&\colhead{(6)}&\colhead{(7)}}
\tabletypesize{\scriptsize}
\startdata
HD 139614 & A7 & 0.3 & 1.80 & 1.77 & 13 & 1,5,11\\
HD 141569 & A0 & 35.7 & 2.33 & 1.94 & 228 & 1,2\\
HD 142666 & A5 & -7.0 & 2.15 & 2.82 & 65 & 1,3\\
HD 169142 & A5 & -0.4 & 1.69 & 1.64 & 48 & 1,5\\
HK Ori & A3 & 14.4 & 3.00 & 4.10 & 60 & 2,4,6\\
KK Oph & A8 & 10.0 & 2.17 & 2.94 & \nodata & 2,6,10\\
RR Tau & A0 & 15.3 & 5.80 & 9.30 & 225 & 2,6,7\\
T Ori & A3 & 56.1 & 3.13 & 4.47 & 147 & 1,2,8\\
VV Ser & B7 & -2.0 & 4.00 & 3.10 & 124 & 1,7,8\\
XY Per & A2 & 2.0 & 1.95 & 1.65 & 224 & 1,3,8\\
DX Cha & A7 & 14.0 & 2.0 & 0.84 & \nodata & 5,9\\
 & & & & & & \\
41 Ari & B8 & 4.0 & \nodata & \nodata & 175 & \nodata \\
$\alpha$ CMa & A0 & -5.5 & \nodata & \nodata & 17 &  \nodata\\
HD 42111 & A3 & 25.3 & \nodata & \nodata & 288 & \nodata\\
\enddata
\tablenotetext{a}{All standard star information taken from SIMBAD; 1=\citet{alecian13}, 2=\citet{mend11}, 3=\citet{db11}, 
4=\citet{hillenbrand92}, 5=\citet{garcia06}, 6=\citet{manoj06}, 7=\citet{cauley15},
8=\citet{mend12}, 9=\citet{grady04}, 10=\citet{finkjank}, 11=\citet{meeus98}}
\end{deluxetable*}
\capstarttrue

\section{LINE PROFILES}
\label{sec:sec3}

The extracted line profiles are shown in \autoref{fig:fig1} and \autoref{fig:fig2}. The rest
velocity of the lines is marked with a black dashed line. Nearby lines that potentially
contaminate the lines of interest are marked with vertical dashed--dotted blue lines.

\begin{figure*}[htbp]
   \centering
   \includegraphics[scale=.80,trim=30mm 20mm 30mm 25mm]{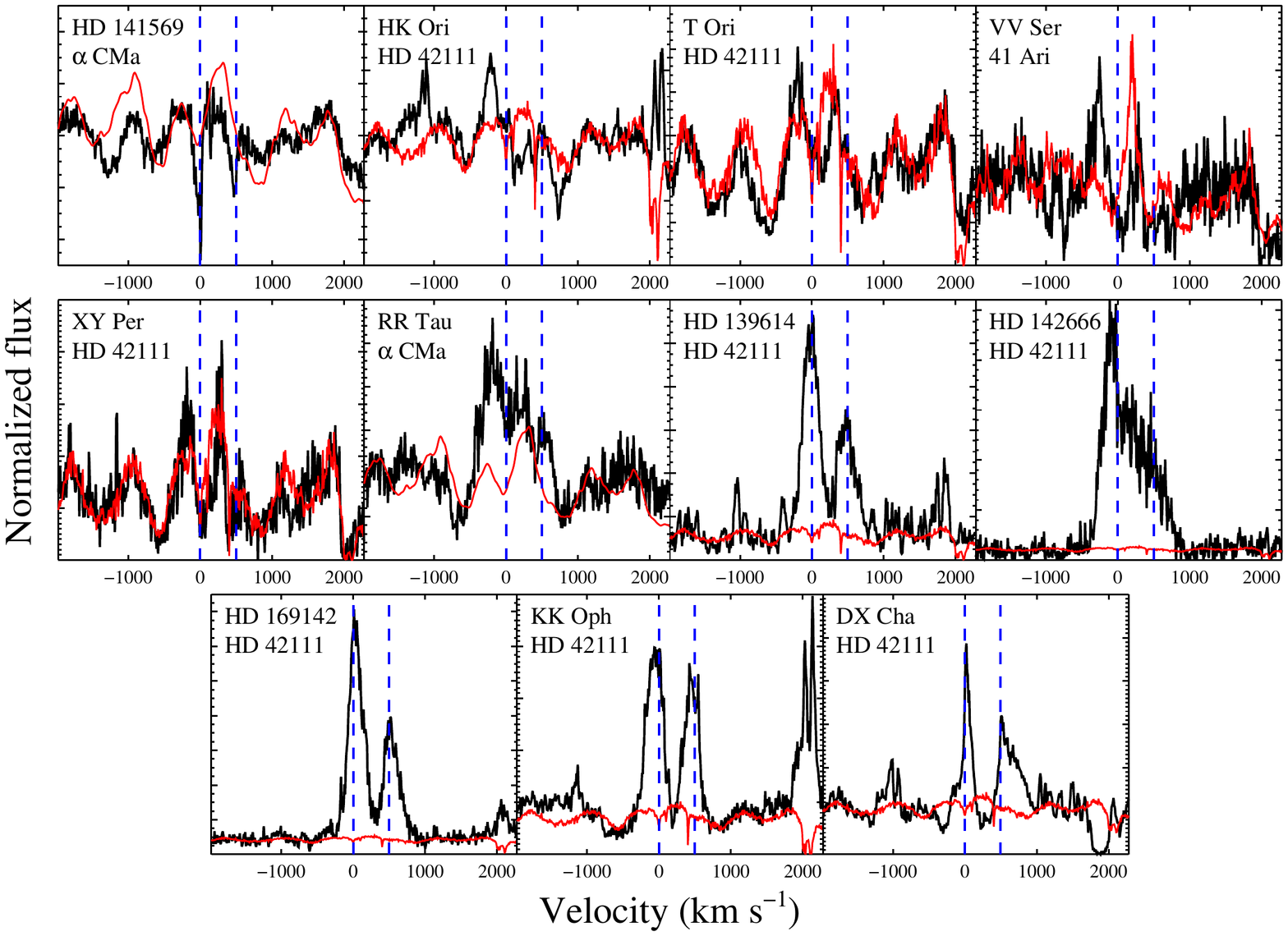} % requires the graphicx package
   \caption{Comparison of the observed \ion{C}{4} lines (black) with a spectroscopic main sequence standard
   of a similar spectral type (red). The standard is scaled to match the photospheric features of the object. The standards shown
   for HD 141569 and RR Tau were rotationally broadened. The rest velocities of the \ion{C}{4} doublet members
   are marked with vertical blue dashed lines.}
   \label{fig:fig2a}
\end{figure*}

Unlike CTTSs, HAEBES have non-negligible photospheric flux at UV wavelengths. To aid in identifying the circumstellar features,
we compare each object spectrum with a main sequence dwarf standard of similar spectral type. These
standards are listed in the last four lines of \autoref{tab:tab2}. The standard spectra were obtained 
from StarCAT\footnote{https://archive.stsci.edu/prepds/starcat/}, a compilation of UV stellar spectra obtained
with STIS \citep{ayres10}. These comparisons are shown for \ion{C}{4} in \autoref{fig:fig2a}.
Although we only show the kinematically important regions of the spectra in \autoref{fig:fig1} and \autoref{fig:fig2},
approximately 10 \AA\hspace{0pt} of spectrum on either side of the lines are compared to
a standard in order to approximately scale and match the standard to the object. For most objects, this allows
unambiguous identification of photospheric features and allows a better estimate of the continuum to 
be made (horizontal red lines in \autoref{fig:fig1} and \autoref{fig:fig2}; see \autoref{tab:tab3}). 

We note that there is a lack of high-resolution UV observations of main sequence, late A-type dwarfs. Thus
for the late A-type stars in our sample we use the A3 standard HD 42111. Due to the strong emission
lines in these objects, the spectral type mismatch does not affect the analysis. 

\capstartfalse           
\begin{deluxetable*}{lcccccccccc}
%\rotate
%\linespacing{1}
\tablecaption{Literature distance, extinction, and accretion rate values \label{tab:tab4}}
\tablewidth{\textwidth}
\tablehead{\colhead{}&\multicolumn{5}{c}{A$_V$ (mags)/Distance (pc)}&\colhead{}&\multicolumn{4}{c}{log$_{10}$($\dot{M}$) ($\Msun$ yr$^{-1}$)}\\
\colhead{Object ID}&\colhead{A13}&\colhead{F15}&\colhead{H04$^b$}&\colhead{G06}&\colhead{V03}&
\colhead{}&\colhead{F15}&\colhead{M11}&\colhead{G06}&\colhead{DB11}\\
\colhead{(1)}&\colhead{(2)}&\colhead{(3)}&\colhead{(4)}&\colhead{(5)}&\colhead{(6)}&\colhead{}&\colhead{(7)}&\colhead{(8)}
&\colhead{(9)}&\colhead{(10)}}
\tabletypesize{\scriptsize}
\startdata
HD 139614 & 0.50 & 0.00 & \nodata & 0.10 & \nodata & & -7.63 & \nodata & -7.99 & \nodata\\
 &  142 & 140 & \nodata & 150 & \nodata & & & & & \\
HD 142666 & 1.60 & 0.50 & \nodata & 0.80 & 1.15 & & $<$-8.38 & -6.73 & -8.00 &-7.77 \\
 & 145  & 145 & \nodata & 116 & \nodata & & & & &\\
HD 169142 & -0.30 & \nodata & \nodata & 0.30 & \nodata & & \nodata & \nodata & -7.40 &\nodata\\
& 145 & \nodata & \nodata & 145 & \nodata & & & & &\\
KK Oph & \nodata & 2.70 & \nodata & 0.44 & 0.90 & & $<$-7.84 & -5.85 & -7.91 &\nodata\\
& \nodata & 279 & \nodata & 160 & \nodata & & & & &\\
RR Tau & \nodata & \nodata & 2.00/3.20 & \nodata & 4.52 & & \nodata & -4.11 & -6.68 & \nodata\\
 & \nodata & \nodata & 800 & \nodata & \nodata & & & & &\\
VV Ser & 5.35 & \nodata & 3.40/5.40 & 3.10 & \nodata & & \nodata & \nodata & -6.34 & -7.49\\
 & 260 & \nodata & 440 & 260 & \nodata & & && &\\
DX Cha & \nodata & 0.00 & \nodata & 0.70 & 0.44 & & -6.68 & \nodata & -7.45 & \nodata\\ 
 & \nodata & 115 & \nodata & 116 & \nodata & & & & &\\
\enddata
\tablenotetext{a}{A13=\citet{alecian13}, F15=\citet{fairlamb15}, H04=\citet{hernandez04}, G06=\citet{garcia06},
 V03=\citet{valenti03}, M11=\citet{mend11}, DB11=\citet{db11}}
\tablenotetext{b}{\citet{hernandez04} calculate $A_V$ using both $R_V$=3.1 (left value) and $R_V$=5.0 (right value).}
\end{deluxetable*}
\capstarttrue

\capstartfalse           
\begin{deluxetable*}{lcccccc}
%\rotate
%\linespacing{1}
\tablecaption{\ion{C}{4} line parameters \label{tab:tab3}}
\tablewidth{\textwidth}
\tablehead{\colhead{}&\colhead{V$^{blue}_{max}$}&\colhead{V$^{red}_{max}$}&\colhead{}&\colhead{$F_{cont}\times10^{14}$}&
\colhead{$F_{CIV}\times10^{14}$$^a$}&\colhead{}\\
\colhead{Object ID}&\colhead{(km s$^{-1}$)}&\colhead{(km s$^{-1}$)}&
\colhead{$F^{blue}_{max}/F^{red}_{max}$}&\colhead{(erg cm$^{-2}$ s$^{-1}$ \AA$^{-1}$)}&\colhead{(erg cm$^{-2}$ s$^{-1}$)}&
\colhead{$L_{CIV}$/$L_\odot$$^b$}\\
\colhead{(1)}&\colhead{(2)}&\colhead{(3)}&\colhead{(4)}&\colhead{(5)}&\colhead{(6)}&\colhead{(7)}}
\tabletypesize{\scriptsize}
\startdata
HD 139614 & -13.5 & -23.2 & 1.71 & 1.23 & 52.7$\pm$0.9 & 3.3-11.0$\times$10$^{-4}$\\
HD 142666 & -105.1 & -14.4 & 2.34 & 0.04 & 20.2$\pm$0.5 & 4.4-61.4$\times$10$^{-4}$\\
HD 169142 & 28.7 & -1.6 & 1.82 & 1.12 & 50.6$\pm$0.9 & 1.6-68.0$\times$10$^{-4}$\\
KK Oph & -50.9 & -55.1 & 1.17 & 0.55 & 14.9$\pm$0.1 & 1.0-234.8$\times$10$^{-3}$\\
RR Tau & -191.2 & -215.4 & 1.22 & 0.25 & 4.2$\pm$0.1 & 0.10-42.9\\
VV Ser & -265.1 & -223.0 & 1.62 & 0.11 & $>$0.4$\pm$0.07 & 1.4-356.0$\times$10$^{-2}$\\
DX Cha & 31.1 & 27.2 & 1.65 & 16.8 & 221.7$\pm$0.8$^c$ & 9.2-49.8$\times$10$^{-4}$\\ 
\enddata
\tablenotetext{a}{Total flux in the line integrated from $-$500 to +900 km s$^{-1}$. Flux uncertainties are derived
from the Poisson errors associated with the reduced exposures.}
\tablenotetext{b}{Ranges for the \ion{C}{4} line luminosities using the median distances and all $A_V$ values from
\autoref{tab:tab4}.}
\tablenotetext{c}{The \ion{C}{4} flux measured here is 1.5 times higher than that measured by \citetalias{ardila}. This
is due to a couple of factors: 1. The choice of continuum is likely slightly different; and 2. We do not ignore the H$_2$ 
emission at $\sim$-200 \kms. The second point is minor compared to the first. Thus the differing values are most likely
the result of the different continuum choice.}
\end{deluxetable*}
\capstarttrue

\subsection{\ion{N}{5} 1238.8,1242.8 \AA}
\label{sec:subsec31}

The \ion{N}{5} doublet in our sample is in emission in 6 out of 11 objects. With the exception of DX Cha, 
none of the emission lines show any clear sign of absorption. The absorption in the blue wing of the
red doublet member of the DX Cha profile is most likely due to a \ion{N}{1} transition since it does not
appear in the blue doublet member. The emission strength in HD 142666, RR Tau, and KK Oph seems to be comparable
for both doublet members, although the true flux level of the red doublet member is obscured by
\ion{N}{1} emission in the RR Tau and KK Oph profiles.

\subsection{\ion{C}{4} 1548.2,1550.8 \AA}
\label{sec:subsec32}

Six of the eleven objects show clear emission in the \ion{C}{4} doublet. Two objects, HD 142666 and
RR Tau show broad emission such that the two distinct members of the doublet are not clearly
identified.  The apparent emission features of T Ori and XY Per are approximately consistent with a
photospheric spectrum of similar spectral type. The profiles of VV Ser and HK Ori show deviations
from the photospheric standard profiles, suggesting a circumstellar component.  However, higher
signal-to-noise observations and a more precise standard comparison are needed to verify the nature
of the difference.  Three objects, HD 139614, HD 169142, and DX Cha, show visual evidence of
asymmetric profile shapes and differing morphologies between doublet members.  HD 141569 is the only
object that shows sharp blue--shifted absorption features.  The absorption extends to $\sim$-150 km
s$^{-1}$, suggestive of a relatively weak outflow. 

%VV Ser may show signs of 
%red--shifted absorption, although the signal--to--noise of the observation makes it difficult
%to locate the potential edges of the red doublet absorption feature. However, when compared with
%a late B--type standard, the local continuum shows a deficit of flux extending to $\sim$400 
%km s$^{-1}$. We note that the red-shifted absorption is present regardless of the spectral standard used in the comparison.
%inverse P-Cygni profiles in \ion{C}{4} were observed for a handful of HAEBES by \citet{grady96}, all of which
%were B and early A--type objects. The maximum absorption velocity for VV Ser of $\sim$400 km s$^{-1}$
%is similar to that observed in this star by \citet{grady96}

\subsubsection{The \ion{C}{4} line luminosity and mass accretion rate}
\label{subsubsec:subsubsec321}

\citet{jk00} were the first to demonstrate a relationship between accretion rate and \ion{C}{4} line luminosity for CTTSs.
They interpreted the excess \ion{C}{4} flux, the flux remaining after magnetic surface activity is taken into account, as
being produced in the accretion flow onto the star. \citet{yang12}, for a sample of 37 CTTSs, confirmed the relationship
between accretion luminosity ($L_{acc}$) and \ion{C}{4} line luminosity. They also showed that CTTSs show greater \ion{C}{4} line
luminosities as a group than WTTSs, strengthening the conclusion that the \ion{C}{4} line flux is produced by mass
accretion. This relationship was further confirmed by \citetalias{ardila} for a
sample of 28 CTTSs, although they point out that, due to the complexity of the physical processes producing the 
\ion{C}{4} flux, along with the intrinsic variability of CTTSs, the exact relationship between $\dot{M}$ and $L_{CIV}$ is 
uncertain. This relationship has not been extended to the higher mass HAEBES, although a weak correlation
was found between $L_{acc}$ and $L_{CIV}$ for the intermediate mass TTSs by \citet{calvet04}.     

We have measured the \ion{C}{4} line fluxes for the HAEBES in our sample that show clear emission lines. This
criteria excludes HD 141569, T Ori, and XY Per from the analysis. We also exclude HK Ori since most of the
flux in the red-ward doublet member is absent or possibly absorbed by the blue-ward doublet member. We include VV Ser
since the emission in the red-ward doublet member is still prominent. 

Estimates of $L_{CIV}$ are very sensitive to $A_V$. Thus the chosen $A_V$ value dominates any random
errors, e.g., the measured flux uncertainties or the formal uncertainties derived for $A_V$ or the object distance.
$A_V$ is also affected by intrinsic accretion and brightness variations of the system, as well as the
choice of total to selective extinction, $R_V$. For most of the objects in our sample, there is a wide range of 
$A_V$ and $\dot{M}$ reported in the literature. Values of $A_V$ can differ based on how they are determined, whether
using photometric colors \citep[e.g.,][]{hernandez04,alecian13,fairlamb15} or spectroscopic template fits 
\citep[e.g.,][]{valenti03}. We have collected these values, as well as distances, from the
recent studies of \citet{alecian13}, \citet{fairlamb15}, \citet{hernandez04}, \citet{garcia06}, \citet{valenti03}, \citet{mend11},
and \citet{db11}. They are given in \autoref{tab:tab4}. We do not give the formal uncertainties from these studies
since, if they are provided, they are typically much smaller than the range of reported values. Instead, we calculate $L_{CIV}$ for the
full range of $A_V$. There is generally better agreement in distance determinations and so we use the median
distance value for each object.

\begin{figure}[htbp]
   \centering
   \includegraphics[scale=.55,trim=50mm 20mm 30mm 40mm]{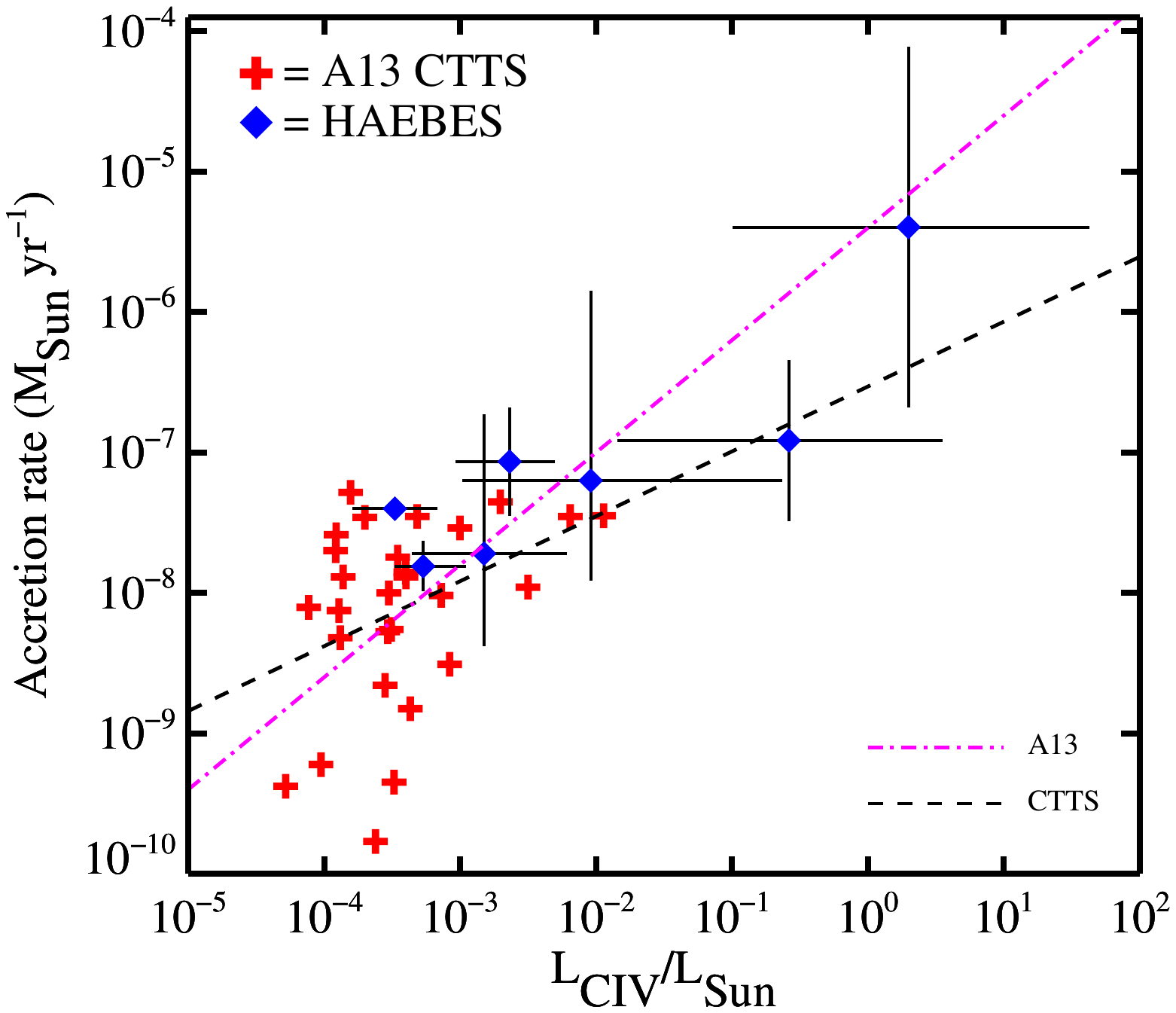} % requires the graphicx package
   \caption{\ion{C}{4} line luminosities versus accretion rate for the extinction--corrected CTTSs from \citetalias{ardila} (red crosses) and the
   HAEBES from this study (blue diamonds). The best--fit relationship for the CTTS sample found by \citetalias{ardila} is
   shown in pink. The best--fit relationship for the \citetalias{ardila} CTTS sample using values of $A_V$ from \citet{herczeg14}
   and \citet{furlan09,furlan11}. The $L_{CIV}$ error bars for the HAEBES represent the minimum and maximum values 
   obtained using each of the literature $A_V$ values from \autoref{tab:tab4}. Error bars for $\dot{M}$ show the maximum
   and minimum literature values from \autoref{tab:tab4}.}
   \label{fig:fig3}
\end{figure}

We choose to use $R_V$=3.1 for the $L_{CIV}$ calculations. There has been some effort by various authors
to determine the appropriate value of $R_V$ for calculating extinction values towards HAEBES. \citet{calvet04}, 
for example, find that anomalous extinction laws that closely resemble the $R_V$=3.1 standard interstellar 
function are better suited for their sample of intermediate mass CTTSs. \citet{hernandez04} find
better agreement between $A_V$ derived from two different colors, $B-V$ and $V-R$, by using $R_V$=5.0.
However, for $A_V$$<$2.0 there is little difference between the $R_V$=3.1 and $R_V$=5.0 cases. Furthermore,
$R_V$=3.1 is adopted by a majority of the studies from which the $A_V$ values are taken; $R_V$=5.0 is
used in \citet{hernandez04} and \cite{alecian13}. Choosing $R_V$ =3.1 thus provides the most consistency 
across studies. We note that using $R_V$=5.0 shifts $L_{CIV}$ to lower values (i.e., to the left in \autoref{fig:fig3}) 
but maintains the trend found using $R_V$=3.1. Given the highly uncertain nature of $L_{CIV}$, regardless of
the chosen value of $R_V$, the adopted $R_V$ does not affect the overall conclusions.

The doublet line flux was calculated by subtracting the continuum flux
from \autoref{tab:tab3} across the line and integrating the flux, centered on the 1548.2 \AA\hspace{0pt} doublet member,
 from -500 to 900 km s$^{-1}$. This velocity range was chosen to be consistent with \citet{ardila}. For VV Ser, only the emission
components of the line profile are included in the integration. We take this to be a lower limit on the total emission. 
The integrated flux value is then corrected for extinction using the $A_V$ values in \autoref{tab:tab4} and an extinction law with $R_V$=3.1.
The line luminosity is then calculated using the median distance in \autoref{tab:tab4}. The final values of $F_{CIV}$ and the
range of $L_{CIV}$ are given in \autoref{tab:tab3}. The accretion rates are not simultaneous with the \ion{C}{4} measurements.

\autoref{fig:fig3} shows $\dot{M}$ plotted against $L_{CIV}$ for the CTTS sample from
\citetalias{ardila} and the HAEBE sample from this study. For the HAEBES, the average value in
log-space of $L_{CIV}$ is shown.  The magenta line is the linear relationship found by
\citetalias{ardila} for their entire CTTS sample. The black dashed in \autoref{fig:fig3} is the
relationship we find for the \citetalias{ardila} CTTSs but using values of $A_V$ from
\citet{herczeg14} and \citet{furlan09,furlan11}. \citet{herczeg14} derive values of $A_V$ by fitting
a large grid of low-extinction spectral templates plus a flat accretion spectrum to their CTTS
sample. \citet{furlan09,furlan11} use photometric colors compared with photospheric template colors,
and an assumption about the underlying excess accretion emission, to calculate $A_V$. The $A_V$
values from \citetalias{ardila} come from a variety of sources but a significant fraction
($\sim$40\%) are taken from \citet{gullbring}. The main difference in the derivation of $A_V$ across
these studies is how the excess accretion and inner disk emission is treated, if it is considered in
detail at all. How the excess emission is determined is likely a larger driver of the differences in
$A_V$ than the difference between photometric and spectroscopic methods. The significantly different
linear fits in \autoref{fig:fig3} demonstrate how the adopted value of $A_V$ can strongly affect the
$L_{CIV}$-$\dot{M}$ relationship.

 Although the number of objects is small, the HAEBES generally show an increase of $L_{CIV}$ with increasing
$\dot{M}$, suggesting a common origin for the line flux in both types of objects. The large spread in $\dot{M}$ and $A_V$
for the HAEBES prevents any meaningful linear relationship from being extended into the HAEBE regime. 
In any case, as shown in \autoref{fig:fig3}, this relationship is very uncertain for the CTTS sample.
Observations of a larger sample of HAEBES, combined with simultaneous determinations of $\dot{M}$ and $A_V$, 
will be beneficial in further constraining the relationship between $\dot{M}$ and $L_{CIV}$ across a broad
stellar mass range. 

\subsection{\ion{Si}{4} 1398.8,1402.8 \AA}
\label{sec:subsec33}

Six objects show clear emission signatures in \ion{Si}{4}. The \ion{Si}{4} line profiles can be contaminated
by nearby CO A-X (5-0) bandhead emission (1393 \AA, -271 km s$^{-1}$) and \ion{O}{4} emission (1401 \AA, +1633 km s$^{-1}$). All objects 
displaying \ion{C}{4} emission, with the exception of VV Ser, show some sign of emission in \ion{Si}{4}. DX Cha
is the only object in the sample that shows a clear blue--shifted absorption signature in \ion{Si}{4}: the emission
profiles are sharply cut off at zero velocity, indicating a strong absorbing wind. DX Cha's \ion{C}{4} line shows
a similar morphology although the red and blue components differ more strongly.

\subsection{\ion{He}{2} 1640.5 \AA}
\label{sec:subsec34}

Compared to the CTTS sample of \citetalias{ardila}, \ion{He}{2} emission is relatively uncommon in our sample: 
only 5 objects (HD 142666, HD 169142, HK Ori, KK Oph, and DX Cha) show clear signs of an emission profile. 
The strong emission lines to both sides of zero velocity
in the line profiles of HK Ori, KK Oph, VV Ser, XY Per, and DX Cha are likely complexes of nearby \ion{Fe}{2} lines. 
The relative absence of \ion{He}{2} emission in our sample, and the presence of strong \ion{Fe}{2} features, contrasts strongly 
with the ubiquitous \ion{He}{2} emission lines, and lack of \ion{Fe}{2} lines, seen in
the CTTS sample from \citetalias{ardila}. This is consistent, however, with the general lack of \ion{He}{1} 5876 \AA\
emission seen in the optical HAEBE sample from \citet{cauley15}. The presence of strong \ion{Fe}{2} features in
the HAEBE UV spectra also reinforces the findings from \citet{cauley15} that \ion{Fe}{1} is largely absent from 
HAEBE spectra due to the intense ionizing flux emitted by HAEBE photospheres. The lack of \ion{He}{2} 
emission in the HAEBE sample will be more fully explored in \autoref{sec:sec5}.

\section{INVESTIGATING THE \ion{C}{4} MORPHOLOGIES}
\label{sec:sec4}

In young stars, \ion{C}{4} line flux can be formed in at least three distinct scenarios: the accretion flow onto the star, hot outflows (e.g., a stellar wind),
or in the transition region/chromosphere. If the emitting region is optically thin, the resulting profile would
be purely in emission; if the optical depth is large enough, the emission line will show signs of self absorption.
However, it can be non-obvious distinguishing between asymmetric emission line profiles that form as a result of
self absorption versus those that form as a result of having multiple emission components.
The \ion{C}{4} doublet is especially useful for identifying potential outflows: 
the velocity separation of the doublet ($\sim$500 \kms) is small enough that optically thick outflows with
terminal velocities near the velocity separation of the doublet should show evidence of absorption in the
\textit{red} wing of the blue doublet member.  

In order to investigate the origin of the observed line profiles, we compare simple,
pure scattering stellar wind model fits (i.e., emission profiles with absorption) with multi-Gaussian fits 
meant to represent emission lines formed in an accretion flow or chromosphere (i.e., pure emission lines with no absorption). 
Simple $\beta$--velocity wind laws, like the one that we employ here, have been shown to be good descriptions of 
the velocity structure in the winds of O--type stars \citep[e.g.][]{lamers76,groenewegen} and multi-Gaussian fits are good approximations
of the pure emission line profiles in CTTSs, which are believed to form in the accretion flow onto the star \citep{ardila}. 

\subsection{Multi--component Gaussian fitting}
\label{subsec:subsec43}

\citetalias{ardila} found that the UV doublet emission line profiles of CTTSs can be accurately described by
fitting two Gaussians or a single Gaussian to each doublet member. The objects requiring a two-Gaussian fit
generally show evidence of a broad component and a narrow component. The broad component is believed
to form along the accretion stream while the narrow component is formed in the post-shock gas at the
stellar surface \citep{calvet98,ardila}. For CTTSs, \citetalias{ardila} argue in favor of these hot gas lines being produced in 
the accretion flow onto the star. They cannot, however, rule out their formation in the stellar transition region, though the
large velocity motions needed for the broad component argues against this interpretation. In this section
we perform multi--component Gaussian fits to the \ion{C}{4} doublets in order to investigate 
the origin of the \ion{C}{4} line emission in HAEBES. Only objects which show pure emission features are
included. This excludes HD 141569, T Ori, and XY Per, which closely match their photospheric templates, and HK Ori
and VV Ser, which show evidence of a non-stellar contribution.

\begin{figure*}[htbp]
   \includegraphics[scale=.75,trim=5mm 10mm 5mm 0mm]{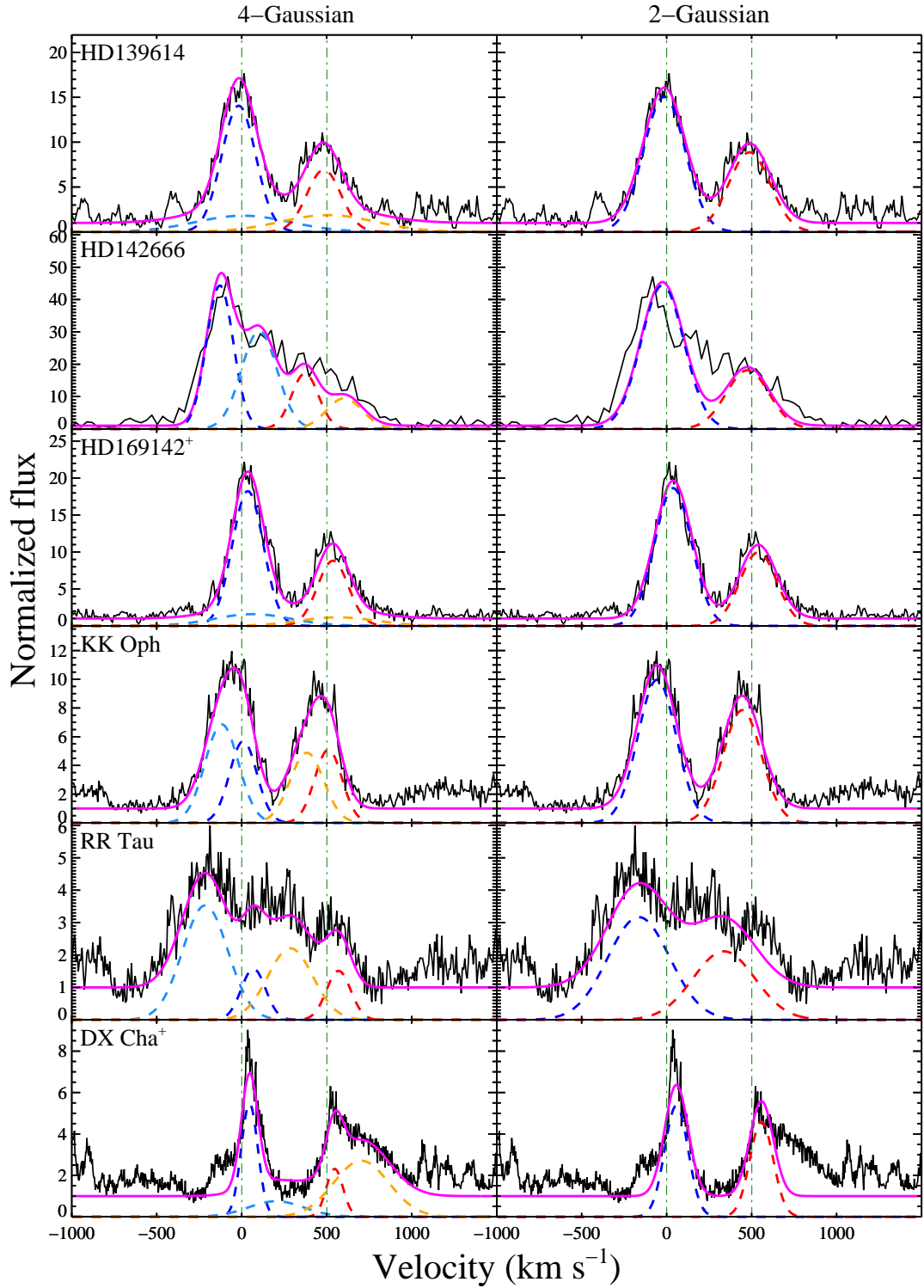} % requires the graphicx package
   \caption{Multi--component Gaussian fit to the \ion{C}{4} emission lines. The left-hand column
   shows the 4-Gaussian fit and the right-hand column shows the 2-Gaussian fit. The data is shown in black and the fit is shown in magenta.
   The individual Gaussians are shown in blue and light blue for the blue doublet member and red and orange for the red 
   doublet member. The rest velocities of the doublet members are marked with vertical green lines. The objects marked with
   a + are the objects we fit using the stellar wind model. Note the large intensity of the broad component for the
   red doublet member of DX Cha (orange dashed line) compared to the broad component of the blue doublet member 
   (light blue dashed line): this is unphysical due to the \ion{C}{4} 1548,1550 \AA\ atomic parameters. For HD 139614,
   HD 169142, and KK Oph, the 2-Gaussian fit is comparable to the 4-Gaussian fit. HD 142666 and RR Tau are
   more accurately described by a 4-Gaussian fit (however, see \autoref{subsec:subsec53}).}
   \label{fig:fig7}
\end{figure*}

We fit each emission doublet with two different combinations of Gaussians: 1. a four--Gaussian fit composed of
a broad and narrow component for each doublet member; 2. a single Gaussian for each doublet
member, i.e., a two--Gaussian fit. The four--Gaussian fit has the following form:

\begin{equation}\label{eq:eq6}
\begin{split}
  F_{tot} = h_1 \exp\Big({-\frac{(v-v_1)^2}{2 \sigma_1^2}}\Big) + h_2 \exp\Big({-\frac{(v-v_2)^2}{2 \sigma_2^2}}\Big) + \\
                h_3 \exp\Big({-\frac{(v-v_1-v_{off})^2}{2 \sigma_1^2}}\Big) + \\
                h_4 \exp\Big({-\frac{(v-v_2-v_{off})^2}{2 \sigma_2^2}}\Big) + 1
\end{split}
\end{equation}  

\noindent where $h$ is the height of the Gaussian, $v$ is the velocity offset, $\sigma$ is the width of the Gaussian in \kms, 
and $v_{off}$ is the velocity separation of the doublet (500.96 \kms).
A continuum of 1.0 is added to the sum of the components. Thus the two broad components have the same
velocity offset, $v_1$, and the same width, $\sigma_1$, and likewise for the narrow components. The two--Gaussian
fit is simply Gaussians 1 and 3 from \autoref{eq:eq6}. We fit for the parameters $h_1$, $v_1$, $\sigma_1$, $h_2$, $v_2$, 
$\sigma_2$, $h_3$, and $h_4$. The two--Gaussian fits involve fitting for $h_1$, $v_1$, $\sigma_1$, and $h_3$. 
We utilize a non--linear least--squares fitting routine based on the Marquardt method to determine
the profile fits \citep[see][]{bevington} and the normalized Poisson uncertainties are used to weight the individual points.
The best--fit parameters for each object showing \ion{C}{4} emission are given in \autoref{tab:tab5}.
The fits and the individual Gaussians are shown in \autoref{fig:fig7}. 

The \ion{C}{4} profiles of HD 139614, HD 169142, and KK Oph are all well described by both a two- and a four--Gaussian fit. 
The spectrum of HD 142666 has very low S/N and we suspect that the significant emission between the
rest velocities of the \ion{C}{4} components is actually a separate \ion{N}{5} emission component 
(see \autoref{sec:subsec31}) and not the result of a narrow and broad \ion{C}{4} component. This is similar to the profile of RR Tau which also
appears to be adequately described by a 4-Gaussian fit. Due to asymmetric appearance of the 
blue and red wings of HD 169142's \ion{C}{4} doublet members, the profile will be investigated using
the wind model in the next section. 

Although the observed doublet components of DX Cha clearly do not have the same shape, we have shown the 
Gaussian fits to the DX Cha profile for illustration. The peak Gaussian fluxes for the broad components of the DX Cha fit (\autoref{tab:tab5})
have a blue--to--red component ratio of 0.30. In an optically thin or effectively thin gas, the blue--to--red component 
ratio of the \ion{C}{4} doublet emission lines is 2.0, as determined by the statistical weights of the atomic levels. 
As the medium becomes more optically thick and the lines thermalize, this ratio approaches 1.0. This is 
evidence that the observed DX Cha line profile is not simply the sum of various emission components
and suggests that the wind interpretation, which will be explored quantitatively in the next section, is a more accurate 
description of the physical process producing the doublet morphology. 

\capstartfalse           
\begin{deluxetable*}{lccccccccc}
%\rotate
%\linespacing{1}
\tablecaption{\ion{C}{4} Gaussian fit parameters \label{tab:tab5}}
\tabletypesize{\normalsize}
\tablewidth{\textwidth}
\tablehead{\colhead{}&\colhead{$h_1^a$}&\colhead{$v_1$}&\colhead{$\sigma_1^b$}&\colhead{$h_2$}&\colhead{$v_2$}&\colhead{$\sigma_2$}&
\colhead{$h_3$}&\colhead{$h_4$}&\colhead{}\\
\colhead{Object}&\colhead{($F_{cont}$)}&\colhead{(km s$^{-1}$)}&\colhead{(km s$^{-1}$)}&\colhead{($F_{cont}$)}&\colhead{(km s$^{-1}$)}&
\colhead{(km s$^{-1}$)}&\colhead{($F_{cont}$)}&\colhead{($F_{cont}$)}&\colhead{$\chi_\nu^2$}\\
\colhead{(1)}&\colhead{(2)}&\colhead{(3)}&\colhead{(4)}&\colhead{(5)}&\colhead{(6)}&\colhead{(7)}&\colhead{(8)}&\colhead{(9)}&\colhead{(10)}}
%\tabletypesize{\normalsize}
\startdata
HD 139614 & 14.1 & -17.7 & 100.4 & 1.8 & 7.0 & 279.2 & 6.9 & 1.9&0.73\\
HD 142666 & 44.3 & -129.1 & 78.9 & 30.3 & 102.5 & 104.8 & 17.2 & 9.5&0.81\\
HD 169142 & 18.2 & 35.7 & 91.3 & 1.6 & 65.0 & 225.0 & 8.8 & 1.2&0.65\\
KK Oph & 5.7 & 8.9 & 80.4 & 6.9 & -117.6 & 99.0 & 5.1 & 4.9&7.52\\
RR Tau & 1.57 & 69.0 & 71.4 & 3.53 & -214.7 & 133.2 & 1.51 & 2.21 & 1.25\\
DX Cha & 5.5 & 43.4 & 49.0 & 0.8 & 199.8 & 162.4 & 2.3 & 2.7&24.7\\
& & & & & & & & & \\
HD 139614 & 15.1 & -13.6 & 100.4 &  \nodata &  \nodata &  \nodata & 8.9 &\nodata&0.83\\
HD 142666 & 44.4 & -25.2 & 78.9 &  \nodata &  \nodata &  \nodata& 18.1 &\nodata&1.13\\
HD 169142 & 18.6 & 38.0 & 91.3 &  \nodata &  \nodata &  \nodata & 10.0 &\nodata&0.71\\
KK Oph & 10.0 & 8.9 & -56.4 &  \nodata & \nodata &  \nodata & 7.89 & \nodata&8.06\\
RR Tau & 3.17 & -163.7 & 182.4 & \nodata & \nodata & \nodata & 2.11 & \nodata & 1.45\\
DX Cha & 5.4 & 43.4 & 54.6 & \nodata&  \nodata &  \nodata & 4.6 & \nodata&99.0\\
\enddata
\tablenotetext{a}{$F_{cont}$ is the chosen continuum flux, indicated in \autoref{fig:fig1} and \autoref{fig:fig2} by the horizontal red 
dashed lines.}
\tablenotetext{b}{The full width at half maximum (FWHM) of the Gaussian is $\sigma$$\times$2$\sqrt{2ln(2)}$}
\end{deluxetable*}
\capstarttrue

\subsection{Stellar wind model}    
\label{subsec:subsec42}

Our pure-scattering model is parametrized and designed to demonstrate optical depth effects
on the profile morphologies in closely spaced doublets (e.g., the UV \ion{C}{4} doublet). We
do not derive physical parameters (e.g., temperature and mass loss rate) of the wind.
We follow a treatment similar to \citet{castor79}. Our model consists of a non--rotating, spherical
star launching a spherically symmetric scattering wind, as suggested for young stars, for example, by \citet{dupree05}. 
The outflowing material is assumed to originate at the stellar surface and is accelerated outward according to \autoref{eq:eq1}:

\begin{equation}\label{eq:eq1}
  v(r) = v_o + (v_\infty-v_o)(1-\frac{R_*}{r})^\beta
\end{equation} 

\noindent where $v_{\infty}$ is the terminal wind velocity, $v_o$ is the initial wind velocity,
$\beta$ is a positive non--zero number, and $r$ is the radial distance from the origin. Here, we assume the wind is
essentially isothermal in that we assume a constant ionization fraction throughout the wind. Under
this assumption the optical depth at each velocity $\tau$ becomes a function of only the density and
$r$. The density at each $r$ can be assigned based on the mass continuity equation assuming a
spherically symmetric outflow with a constant mass loss rate:

\begin{equation}\label{eq:eq2}
  \rho(r) = \frac{\dot{M}}{4\pi r^2 v(r)}
\end{equation}  

\noindent where $\dot{M}$ is the mass loss rate in the wind. For our purposes, the specific value of
$\dot{M}$ is irrelevant since we only care about the fraction of the original material along the
line of sight to the star at each $r$ and we parameterize the wind in terms of the total optical
depth. 

\begin{figure}[htbp]
   \includegraphics[scale=.60,trim=55mm 40mm 30mm 50mm]{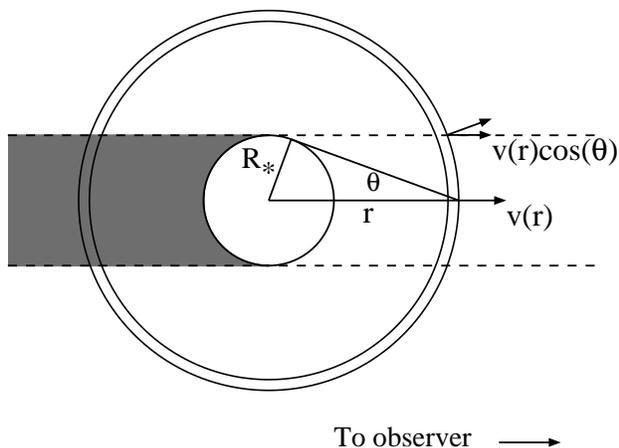} % requires the graphicx package
   \caption{Geometry of the stellar wind model. The shaded gray region illustrates the scattering region
   that does not contribute to the emission profile since these photons are absorbed by the star. The observer is to the right.}
   \label{fig:fig5}
\end{figure}

The integration along the line of sight through the wind is done through a series of equally spaced
shells in velocity space. The initial and final velocities of a shell each define a unique radial
distance from the star according to \autoref{eq:eq1}. We also extend the range of absorbing velocities on
either side of each shell to plus and minus half of the chosen velocity step. This acts as a natural
line width, which we designate $\delta$$v$, since one is not explicitly included in the model. We define the optical depth at each
absorbing velocity shell $v$ as

\begin{equation}\label{eq:eq3}
  \tau(r_v) = \frac{\rho_r \Delta r}{\sum\limits_{i} \rho_i \Delta r_i} 
\end{equation}

\noindent where $r_v$ is the center of the annulus at velocity $v$. Thus each annulus of physical
width $\Delta$$r$ contains a fraction of the total optical depth, which is the optical depth at line
center for the annulus at $r_v$, specified by \autoref{eq:eq3}. 
For each annulus, the amount of absorbed intensity is then equal to $I_r$$e^{-\tau(r_v)}$ where
$I_r$ is the intensity incident upon the annulus. The absorbing velocities range from the maximum shell velocity
down to $v_f$=$v_r$cos$\theta$-$\delta$$v$ where $\theta$=sin$^{-1}$($R_*$/$r$). Thus as the wind moves
farther from the star the range of absorbing velocities becomes smaller. The geometry of the model
is shown in \autoref{fig:fig5}.

We also assume single, isotropic, and completely redistributive scattering \citep{lamers99}. Since the scattering is
isotropic, the amount of scattered intensity in any annulus has to be equally redistributed in emission across velocities that
range from -$v$($r$)-$\delta$$v$ to $v$($r$)$+\delta$$v$.
Emission from velocities being blocked by the star (i.e., velocities of receding material immediately behind
the star) is then set to zero since these photons hit the stellar surface
and do not propagate back to the observer. The final line profile is the sum of the
absorption profile and the emission profiles at each annulus. 

\begin{figure*}[htbp]
   \includegraphics[scale=.76,trim=20mm 65mm 10mm 40mm]{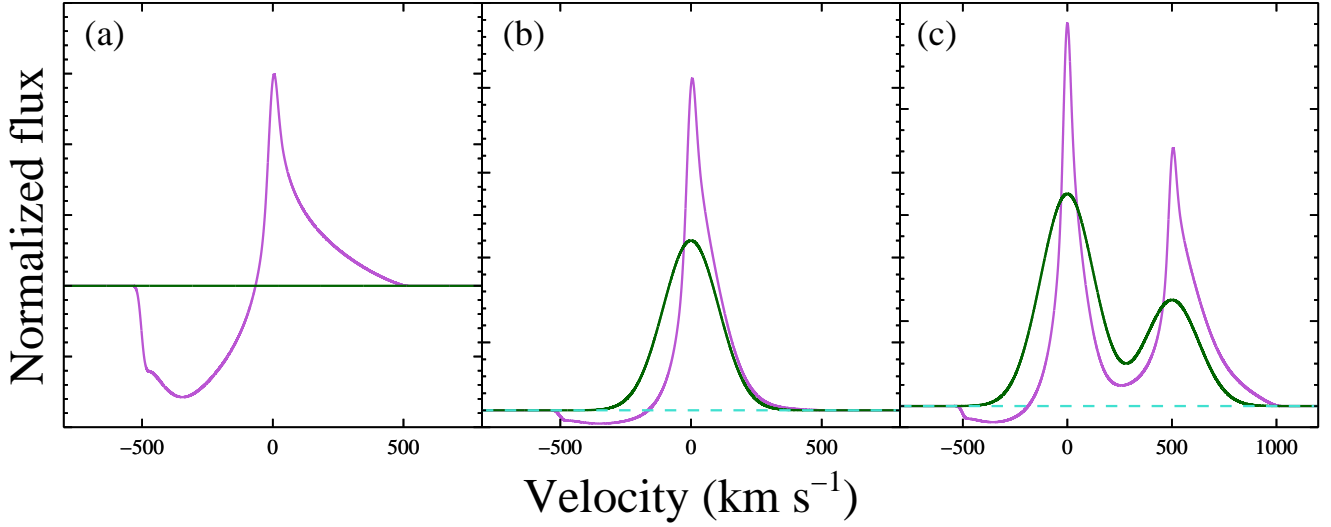} % requires the graphicx package
   \caption{Examples of the stellar wind model for different input spectra. The input spectrum is shown in green,
   the model output in purple, and in panels (b) and (c) the local continuum is shown in turquoise. Each model is calculated using the
   wind parameters $v_{term}$=500 km s$^{-1}$, $\tau$=200., and $\beta$=1.0. Panel (a) shows the case
   of a pure continuum input spectrum; panel (b) shows a single, strong Gaussian emission line as the input
   spectrum; panel (c) shows the case of a strong Gaussian emission doublet as the input spectrum. The resulting
   P--Cygni morphology of the panel (a) spectrum is obvious. In panel (c), the narrower width of the blue--ward
   doublet member can be seen, a result of the red--ward doublet member absorbing the flux in the red velocity
   wing of the blue--ward doublet member.}
   \label{fig:fig8}
\end{figure*}

Examples of the model for various input spectra
are shown in \autoref{fig:fig8}. Panel (a) shows a P-Cygni profile with a flat continuum as the input spectrum,
similar to the profiles computed in the atlas by \citet{castor79}. Panel (b) shows the effect of having a single
Gaussian as the input spectrum, such as proposed by \citet{dupree05,dupree14} for TW Hya. We note that
the result is not a Gaussian but rather a P-Cygni profile. For significant absorption to be present in the blue
wing while simultaneously preserving the Gaussian shape of the red wing of the input spectrum, few of the absorbed 
photons can be scattered back into the line of sight and the blue-wing continuum must be negligible so
as not to reveal the sub-continuum absorption caused by the optically thick wind. If these conditions do not
hold, the resulting line profile will not be Gaussian. Finally, panel (c) shows
the resulting asymmetric line shapes of an optically thick wind in a doublet with a velocity separation
identical to \ion{C}{4}: the red doublet member absorbs some of the flux that has been redistributed in the
red wing of the blue doublet member.

The input spectrum for the fitting routine is a two--component Gaussian with a peak flux ratio of 2.0, which represents doublet emission
from, for example, the base of an accretion flow or a chromosphere. Each component has
the same full width at half maximum (FWHM) and they are centered at their respective wavelengths. 
Both the Gaussian FWHM and the ratio of the peak fluxes are allowed to vary
in the fit. We also allow the total optical depth in the wind, $\tau$, to vary. The terminal velocity and velocity 
parameter $\beta$ are kept fixed at $v_{\infty}$=500 \kms and $\beta$=1.0. We note that changing these
fixed values to other reasonable numbers does not change the general results for most choices of 
$v_{\infty}$ and $\beta$. The same fitting routine mentioned in \autoref{subsec:subsec43} is used for the stellar wind fits. 

\begin{figure*}[htbp]
   \includegraphics[scale=.80,trim=20mm 35mm 40mm 50mm]{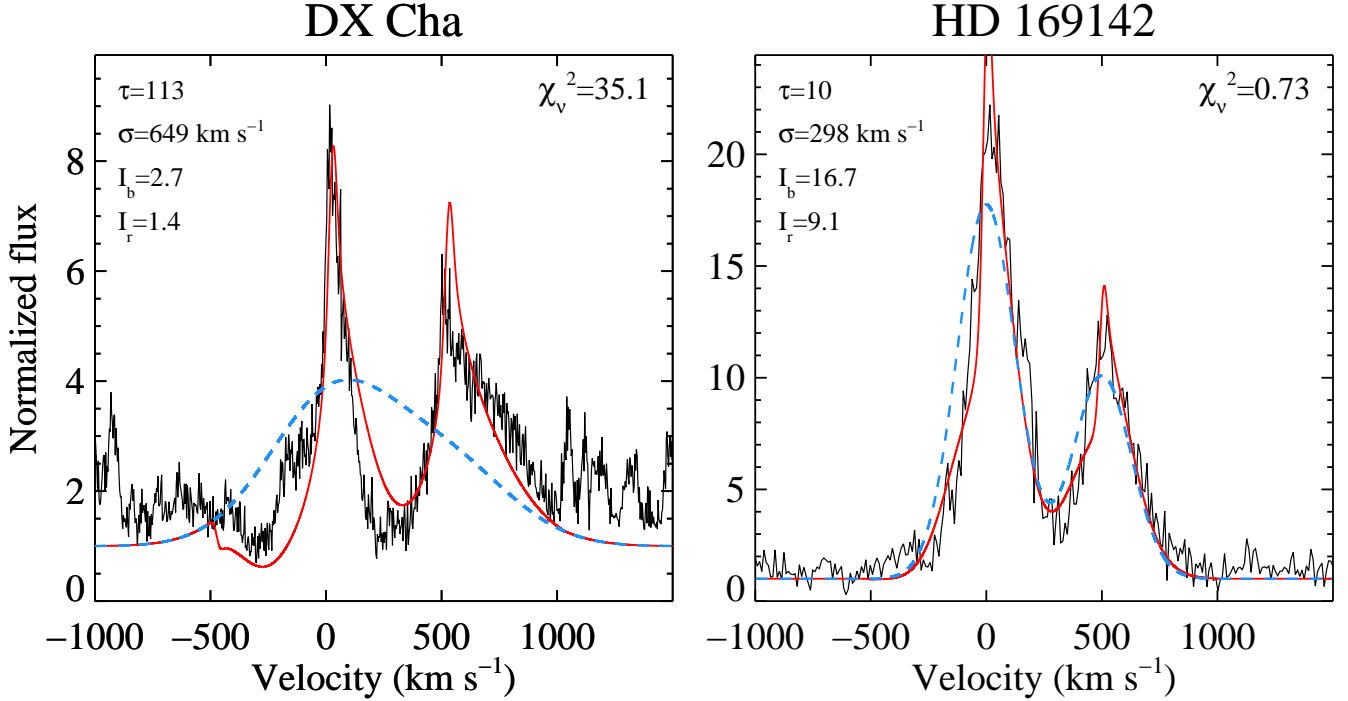} % requires the graphicx package
   \caption{Stellar wind model fits to the objects showing evidence of profile asymmetry. The observed normalized
   profile is shown in black; the best--fit input doublet profile is over plotted in blue; the best-fit wind model is
   shown in red. The total optical depth in the wind is shown in the upper--left corner of each panel, along with
   the best--fit input doublet emission line parameters: $I_b$ is the normalized peak intensity of blue doublet member,
   $I_r$ is the normalized peak intensity of the red doublet member, and $\sigma$ is the width of the input Gaussian. 
   HD 169142 requires a marginal amount of outflowing material to explain the profile shape. 
   DX Cha requires a large $\tau$ to reproduce the profile morphology, indicating the presence of a strong outflow. 
   The emission bump at $\sim$-185 \kms is H$_2$ emission and is ignored in the fit.}
   \label{fig:fig6}
\end{figure*}

The results of the stellar wind model applied to DX Cha and HD 169142 are shown in
\autoref{fig:fig6}. It is immediately clear that the wind model favors a large optical depth for DX Cha and a
moderate $\tau$ for HD 169142. Although
the fitting program favors a small amount of optical depth for HD 169142, the differences between the best--fit model and the observed
profile are well within the measured flux uncertainties. In addition, $\chi_\nu^2$ is worse
for the wind than for either of the Gaussian fits. Thus we find
little evidence for a hot wind in \ion{C}{4} for HD 169142 and favor the multiple emission
components explored in \autoref{subsec:subsec43}. 

DX Cha is the only object in the sample that clearly shows evidence of a strong outflow in the
\ion{C}{4} doublet: the stellar wind model requires a total optical depth of $\tau$=113 to reproduce
the shape of the profile. While the model does a fair job of reproducing the sharp cutoff at zero velocity 
and the narrow shape of the blue doublet component, the pure scattering treatment does a poor
job of reproducing the red--shifted emission in the red component. This is not surprising as the
model assumptions of spherical symmetry and a single emission component are likely incorrect. 
The stellar wind model clearly demonstrates the asymmetry of the red wing of the blue and red 
doublet components in the presence of a strong outflow. In addition, the multi-Gaussian fit to the DX Cha 
profile results in unphysical parameters for the emission lines, further suggesting that the profile 
morphology is the result of a hot outflow.

\section{DISCUSSION}
\label{sec:sec5}

Although the sample of HAEBES presented here is small, the UV line profile morphologies and characteristics
are diverse. Below we discuss some of the important features of the sample, how they relate to
mass accretion and outflows, and how these characteristics compare to those of the CTTS sample
from \citetalias{ardila}.

\subsection{The general absence of hot optically thick outflows}
\label{subsec:subsec51}

The only object in our sample that shows clear evidence of an optically thick outflow in \ion{C}{4}
is DX Cha.  The \ion{C}{4} profile of HD 141569 shows signs of blue--shifted absorption but the
absorption is very narrow and relatively weak. With the exceptions of DX Cha, HD 139614, and KK Oph,
for which we do not have optical or \ion{He}{1} 10830 \AA\hspace{0pt} data, our sample also shows no
direct evidence of optically thick outflows in their optical or \ion{He}{1} 10830 \AA\hspace{0pt}
line profiles. Thus most of these objects not only show no evidence for hot outflows in the UV lines
examined here, wind signatures are also absent from other wavelength regions of their spectra. The
lack of observed outflows may be the result of HAEBES having relatively smaller magnetospheres
compared to CTTSs. This possibility is discussed further in the next section. Regardless, it appears
that hot, optically thick outflows are not common among our sample and, based on the
relatively small percentage of objects showing optical (37\%) or \ion{He}{1} 10830 \AA\ (40\%)
blue-shifted absorption \citep[e.g.,][]{cauley14,cauley15}, possibly among HAEBES in general. 
However, the lack of hot winds should be examined with a larger and more diverse sample of HAEBES.

The strong outflow seen in \ion{C}{4} for DX Cha (also known as HD 104237A) may be related to the presence of 
its short period T Tauri star companion \citep{feigelson03,bohm04}. It is likely part of a quintuple system
with two or more T Tauri star companions orbiting DX Cha and its immediate companion (HD 104237B) at a
distance of $\sim$1500 AU \citep{feigelson03}. Interactions between DX Cha and its close companion could
drive the wind observed in \ion{C}{4}. This would explain why we observe a wind only for DX Cha since none
of the other objects in our sample have confirmed companions with such close separations. Followup observations
should be conducted to measure the \ion{C}{4} flux as function of orbital phase in order to investigate the
effect of close companions on HAEBE \ion{C}{4} flux and profile morphology.

\subsection{A relative paucity of \ion{C}{4} lines}
\label{subsec:subsec52}

We suggested in \citet{cauley14,cauley15} that HAEBES have smaller magnetospheres than CTTSs and
that the relatively smaller disk truncation radii in HAEBES result in less of the necessary energy
from the accretion flow being available to launch outflows from near the stellar surface. Although
neither HAEBES nor CTTSs, in general, show evidence for strong outflows in the hot UV lines, the
paucity of \ion{C}{4} and \ion{He}{2} emission in our sample compared to the CTTS sample from
\citetalias{ardila} suggests that the smaller relative size of HAEBE magnetospheres and their
apparent lack of strong magnetic fields may play a role in the lack of emission in the hot lines. We
note that the free--fall velocities from, for example, $\sim$1.2 $R_*$ for the objects in our sample
are similar to the free--fall velocities from 2.0--2.5 $R_*$ for a typical CTTS of mass 0.5 $\Msun$
and radius 1.5 $\Rsun$. Then for a similar(larger) accretion rate, similar(greater) amounts of
kinetic energy should be dissipated in the accretion shock. This would suggest that if hot outflows
are a product of energy dissipation from the accretion flow that the hot gas lines in emission
should be present in the spectra of accreting HAEBES. Thus the fact that the UV lines are observed
in emission in only $\sim$50\% ($\sim$60\% if the profiles of VV Ser and
HK Ori are included) of our sample, when all of our objects have measured accretion rates, suggests
that the lack of strong emission in the hot UV lines is not a result of these objects not accreting
material from their disks. Furthermore, both XY Per and T Ori, which show no evidence of \ion{C}{4}
emission, show clear inverse P-Cygni profiles in their optical spectra. 

One possible explanation for the relative lack of \ion{C}{4} and \ion{He}{2} emission in our sample 
is that the temperature of the accretion shock in HAEBES is too low to produce
these species. Temperatures below $\sim$10$^5$ K may be insufficient to produce enough observable
\ion{C}{4} flux. If we consider the strong shock regime, i.e., high Mach number, we can estimate the
shock temperature using equation (9) from \citet{calvet98}. For typical HAEBE parameters of
$M_*$=2.0 $M_\sun$ and $R_*$=2.0 $R_\sun$, shock temperatures are limited to 10$^5$ K if material
accretes from $r$$<$1.2 $R_*$. A typical infall radius inferred from the red-shifted absorption
velocities in \citet{cauley15} is $R_{in}$$\sim$1.1 $R_*$. This is significantly smaller than that
for the CTTS sample from \citet{fischer08}, which we calculate to be $R_{in}$$\sim$2.7 $R_*$. In
reality, in-falling material likely originates from farther out in the disk. Projection effects may
result in observed maximum red-shifted velocities that are smaller than the true maximum infall
velocity of the material \citep{fischer08}. In general, however, the smaller values of
$V_{red}$/$V_{esc}$ measured in \citet{cauley14,cauley15} are compatible with the idea that the
infalling material for HAEBES may not be able to heat the post-shock material to temperatures
capable of producing \ion{C}{4} emission. 

For the stellar parameters typical of our sample and using $R_{in}$=1.1 $R_*$, the shock temperature
is only $\sim$3.5$\times$10$^4$ K compared to $\sim$8.5$\times$10$^5$ K for the CTTS sample. Using
the stellar parameters from \autoref{tab:tab2}, we calculate a typical corotation radius of
$R_{co}$$\sim$1.6 $R_*$ for the rapid rotators\footnote{While there are four objects with low
$v$sin$i$ values, HD 142666 is the probably the only object that is truly a slow rotator, the other
low $v$sin$i$ values being the result of viewing the stars nearly pole-on \citep[see][]{meeus98}.}
($v$sin$i$$>$100 km s$^{-1}$) in our sample. This is roughly consistent with the the fact that
material accreting along magnetic field lines must originate at distances \textit{less} than the
corotation radius \citep[e.g.,][]{shu94}. However, if the true infall velocities are higher than the
observed red-shifted velocities by only $\sim$20\%, then the infall distance condition ($<$1.2
$R_*$) required for shock temperatures below 10$^5$ K is no longer met for any of the HAEBES from
\citet{cauley15} with red-shifted absorption profiles. The effect of observing the accretion stream
projected against the star could easily produce a difference of this magnitude. While it is
plausible that the shock temperature of infalling material onto HAEBES could be lower than that for
CTTSs, it is unclear exactly how large this difference is and if it is enough to explain the
relative lack of \ion{C}{4} and \ion{He}{2} emission seen here. Although this idea deserves further
investigation, we tentatively suggest that the evidence of smaller magnetospheres in HAEBES is
further supported by the \ion{C}{4} analysis presented here and, as a result of the lower shock
temperatures produced by the infalling gas, \ion{C}{4} production is stunted compared to CTTSs.

\subsection{The origin of the \ion{C}{4} emission lines}
\label{subsec:subsec53}

\citet{ardila} established that most of the \ion{C}{4} emission observed from CTTSs requires
a 4-component Gaussian to describe the line shape (a broad and narrow component for each
doublet member) and is consistent with formation in an accretion flow, although a contribution 
from the stellar transition region cannot be ruled out. As we showed in \autoref{subsec:subsec43}, the
\ion{C}{4} emission lines of HD 139614, HD 169142, and KK Oph are well described
by a two component Gaussian. Of the pure emission profiles, HD 142666 is the only one
that requires a 4-Gaussian fit. However, as we discussed in \autoref{sec:subsec32}, the
extra emission between the \ion{C}{4} doublet members may be due to \ion{N}{5}. In any
case, one commonality among the \ion{C}{4} emission line profiles of HD 139614, HD 142666,
KK Oph, and RR Tau is that the peak flux is blue-shifted. HD 169142 shows a positive velocity for the 
peak flux in the blue doublet member. 

In general, the blue-shifted velocities measured for the majority of the emission lines, and the
hint of wind absorption in HD 169142, suggests that the \ion{C}{4} lines are being produced in
optically thin outflowing material. If the \ion{C}{4} lines in our sample are the result of wind
emission and a single emission component is required to fit the line morphology, this suggests that
any mass accretion onto these objects is not directly responsible for the emission since emission
produced by in falling material would result in red-shifted emission or emission centered close to
zero velocity. Although this may seem to contradict the findings from
\autoref{subsubsec:subsubsec321}, we are not inferring that mass accretion is not powering the UV
lines but rather that the lines do not form directly in the accretion flows, e.g., along an
in-falling accretion stream. Some of the accretion energy could be used to heat the atmosphere of
the star and power the outflows, i.e., an accretion driven wind \citep{matt05}. As the 
Gaussian fitting demonstrated, the \ion{C}{4} lines also seem to be pure emission
profiles. In other words, they do not seem to be subject to any significant amount of absorption in
the flow. Thus we suggest that the \ion{C}{4} emission line profiles in our sample are formed in
weak, optically thin outflows, perhaps in an expanding stellar chromosphere or corona as first
proposed by \citet{catala84} and \citet{catala88} for AB Aur, that are likely accretion powered to some degree.
This should be further tested with a larger sample of high signal-to-noise \ion{C}{4} HAEBE line
profiles.   

\section{SUMMARY AND CONCLUSIONS}
\label{sec:sec6}

We have presented new high resolution UV spectra for a small sample of HAEBES and analyzed the line
morphologies of the \ion{C}{4} 1548,1550 \AA\hspace{0pt} doublet for evidence of hot ($\sim$10$^5$ K)
winds. Although a larger sample of HAEBES should be examined to better understand these lines and
their formation in HAEBES as a whole, we tentatively conclude the following:

\begin{itemize}

\item The $L_{CIV}$ vs. $\dot{M}$ relationship seems to extend to the HAEBE mass regime, suggesting
a common origin for both CTTSs and HAEBES. A larger sample size of HAEBES is needed to confirm this.
Differing values of $A_V$ can have a drastic effect on the $L_{CIV}$-$\dot{M}$ relationship and simultaneous 
measurements of $L_{CIV}$ and $\dot{M}$ are desirable.

\item With the exception of RR Tau and HD 142666, the \ion{C}{4} emission lines in our sample are
well described by a single Gaussian component. This contrasts with the \ion{C}{4} lines seen in
CTTSs which require both a broad and narrow component for each doublet member. This fact, combined
with the blue-shifted velocities observed for the peak fluxes of the pure emission lines, suggests
that these line profiles in our sample are formed in weak outflows near the stellar surface and not
in accretion flows onto the star. These outflows are likely accretion powered since all of these
objects show signs of accretion in the optical. 

\item Only one object in our sample, DX Cha, shows strong evidence of an optically thick outflow in
\ion{C}{4}. This is demonstrated with a simple wind scattering model.  The existence of a strong
wind may be due to the presence of a low mass companion with an orbital period of $\sim$20 days
\citep{bohm04}. Although blue--shifted absorption has been observed in \ion{C}{4} in a few HAEBES
\citep[e.g.,][]{grady96}, hot, optically thick outflows seem to be uncommon for HAEBES that
also do not show evidence of outflows in the optical or at \ion{He}{1} 10830 \AA. Due to the
relatively smaller percentage of HAEBES with blue-shifted absorption in their optical and
\ion{He}{1} 10830 \AA\ lines, the lack of hot outflows observed here suggests they are also
relatively uncommon for HAEBES in general. However, the statistics on the occurrence rate of winds
in \ion{C}{4} are currently poor. An effort should be made to increase the number of objects with
measured \ion{C}{4} profiles so that the morphologies and line characteristics can be adequately
studied in a manner similar to \citetalias{ardila}. 

\end{itemize}

{\bf Acknowledgments:} We are grateful to the referee for their comments which helped improve
the quality of the paper. We acknowledge funding from grant HST-GO-12996.001 from STSci and from grant
NNX13AF09G through the NASA ADAP program. This research has made use of the SIMBAD database
operated at CDS, Strabourg (France) and NASA's Astrophysics Data System.

\end{document}